\pdfoutput=1
\documentclass[10pt]{article}
\usepackage[letterpaper]{geometry}
\usepackage{hicss51}
\usepackage{times}
\usepackage[none]{hyphenat}
\usepackage{url}
\usepackage{latexsym}
\usepackage{indentfirst}
\usepackage{graphicx}
\graphicspath{{images/}}

\PassOptionsToPackage{hyphens}{url}
\usepackage{url}
\usepackage{breakurl}
\usepackage[breaklinks]{hyperref}

\usepackage{cleveref}
\usepackage{multirow}
\usepackage{todonotes}

\usepackage{xcolor}
\usepackage{adjustbox}

\title{``Healthy surveillance'': Designing a concept for privacy-preserving mask recognition AI in the age of pandemics}

\author{Niklas Kühl\\
Karlsruhe Institute\\ of Technology (KIT)\\
IBM Germany \\
\underline{niklas.kuehl@kit.edu}\And

Dominik Martin\\
Karlsruhe Institute\\ of Technology (KIT)\\
\underline{dominik.martin@kit.edu}\And

Clemens Wolff\\
Karlsruhe Institute\\ of Technology (KIT)\\
\underline{clemens.wolff@kit.edu}\And 

Melanie Volkamer\\
Karlsruhe Institute\\ of Technology (KIT)\\
\underline{melanie.volkamer@kit.edu}
}

\date{}

\begin{document}
\maketitle
\begin{abstract}
The obligation to wear masks in times of pandemics reduces the risk of spreading viruses. In case of the COVID-19 pandemic in 2020, many governments recommended or even obligated their citizens to wear masks as an effective countermeasure. In order to continuously monitor the compliance of this policy measure in public spaces like restaurants or tram stations by public authorities, one scalable and automatable option depicts the application of surveillance systems, i.e., CCTV. However, large-scale monitoring of mask recognition does not only require a well-performing Artificial Intelligence, but also ensure that no privacy issues are introduced, as surveillance is a deterrent for citizens and regulations like General Data Protection Regulation (GDPR) demand strict regulations of such personal data. In this work, we show how a privacy-preserving mask recognition artifact could look like, demonstrate different options for implementation and evaluate performances. Our conceptual deep-learning based Artificial Intelligence is able to achieve detection performances between 95\% and 99\% in a privacy-friendly setting. On that basis, we elaborate on the trade-off between the level of privacy preservation and Artificial Intelligence performance, i.e. the ``price of privacy''.
\end{abstract}

\section{Introduction}
\label{sec:introduction}

The COVID-19 disease has evolved into a global pandemic at the beginning of 2020. In order to fight the spread of the virus, different measures, so-called non-pharmaceutical interventions (NPIs), were taken. One of these measures, which a large share of countries adopted, was the recommendation to wear masks in public spaces \cite{aljazeera2020}. For our work, we define public spaces as any inside or outside area which is generally accessible to people, including publicly operated areas like libraries, tram stations, or public authority buildings, but also privately owned spaces, including restaurants or stores. While it is discussed controversially how high the impact of such a policy is, Greenhalgh et al. (2020) \cite{greenhalgh2020face} conclude that it does help in reducing the viral transmission. For instance, the Czech Republic was one of the first European countries to enforce mask wearing and first analyses point towards that NPI having a major impact on the low number of COVID-19 cases \cite{howard2020face}.

However, it remains of interest whether citizens comply with the directive to wear masks for various reasons---including hesitation to get more evidence whether the mask policy contributes successfully to contain the pandemic, for enforcement reasons and/or to fine none-compliance. While manual inspections, e.g., at the entrance of restaurants or stores, are a possibility, these actions require manual labor, do not scale well and are difficult to enforce on larger spaces. In order to allow for an automated examination of the compliance, one could imagine to use surveillance solutions in combination with Artificial Intelligence (AI) \cite{kuhl2019machine}. 

While this solution entails many upsides, e.g., scalability and automation capabilities, it needs to be in-line with the privacy regulations such as the General Data Protection Regulation (GDPR) and it needs to be understood by  citizens to trust and accept the approach:  
The protection of personal data, e.g., video streams revealing individuals' faces, is regionally required by legal regulations, such as the GDPR in the European Union \cite{GDPR2016}\footnote{This also includes where and how long personal data is stored, which we will discuss at a later point when we explore different implementation options in \Cref{sec:artifactdesign}.}. As numerous studies show, people feel more insecure when their personal steps are highly traceable and they compromise on their privacy while being recorded \cite{slobogin2002public,taylor2010spy,stuart2017beyond}.

Therefore, we propose an end-to-end AI-based surveillance artifact which ensures both (a) privacy and (b) high performance of mask recognition. The proposed solution could be used in diverse application scenarios, for instance to contribute to a rigorous reporting of mask coverage used for research purposes. For instance, it could be implemented as a monitoring capability for impact analyses. As simulation models of pandemics depend on a multitude of parameters as input, a precise assessment of the mask coverage would prove helpful and could improve predictions of pandemic developments, e.g. to analyze the effectiveness of NPIs \cite{baiercovid}. On another note, it could be implemented for private store owners, e.g. to ensure a rigorous reporting of mask coverage to authorities in their privately-owned spaces. 
Note, one aspect we do not regard in this work are countermeasures if citizens or customers do not comply with wearing a mask, as we solely propose a monitoring option at this stage. 

With our research, we contribute to the body of knowledge with three core aspects: First, we develop a novel artifact which can be utilized to allow for a privacy-preserving monitoring of mask coverage during a pandemic. Second, we evaluate different design choices on how to build the artifact and elaborate on their performances, strengths and weaknesses. Third and finally, we theorize on the trade-off between privacy preservation and AI performance---as AI performance decreases with increased privacy preservation and vice-versa. 


As an overall research design, we choose Design Science Research (DSR) and base our approach on Hevner and Chaterjee (2010) \cite{hevner2010}. The authors suggest that a DSR project should cover at least three cycles of investigation, a relevance cycle (targeting the practical problem, see \Cref{sec:relevancecycle}), a rigor cycle (elaborating on the existing knowledge base, see \Cref{sec:rigorcycle}), and one or multiple design cycles (building and evaluating the research artifact, see \Cref{sec:artifactdesign,sec:artifactevaluation}). We finish our work with a discussion on the broader impact (\Cref{sec:discussion}) as well as a summarizing conclusion (\Cref{sec:conclusion}).

\section{Relevance Cycle: Defining the Problem}
\label{sec:relevancecycle}

Recent studies stress: ``If correctly used [...] face masks [...] can contribute to reducing viral transmission'' \cite[p. 1]{royal2020}. 
However, more evidence and more insides on the level of influence are missing. To get these insides, it is necessary to monitor the mask coverage rate (and put the results in relation with other factors). But, especially in public spaces, a monitoring of the mask coverage proves difficult by manual means. Therefore, first countries experiment with automated, AI-based closed-circuit television camera (CCTV) solutions for monitoring. For instance, France is reportedly testing AI-based surveillance tools to check whether people are wearing masks on public transport. To allow for this functionality, the country updated their existing surveillance setup with additional software to allow for monitoring mask-wearing, but also social distancing \cite{bbc2020}. However, we lack information on where the analyses was performed, whether and how this complies with the regulatory requirements of the national GDPR implementation---as well as details on the utilized AI technology.

While AI-based surveillance would in fact be a technically feasible solution to monitor mask recognition, any type of CCTV typically raises privacy concerns \cite{nguyen2011situating}. The mass surveillance in the wake of China's social credit system raised extensive concerns \cite{qiang2019road}. Individual examples of misuse of CCTV exist as well, for instance, investigations were launched after a museum guard used CCTV to spy on Angela Merkel's private apartment \cite{Cavailaro2007}. 

To account for privacy in surveillance we, therefore, aim to design artifacts ensuring both, privacy-preservation and a high-performing recognition of masks. To do so, we lay out different options as depicted within \Cref{sec:artifactdesign}. Initially though, we are interested in existing work in the vicinity of our proposed approach.

\section{Rigor Cycle: Related Work}
\label{sec:rigorcycle}

In this section, we ensure rigorous research by elaborating on related work. To that end, we specifically focus on privacy protection within the European Union (EU) and algorithms to preserve privacy in video-based surveillance systems.

\subsection{Data Protection and Privacy}

In 2018, the EU implemented the GDPR. Among its primary objectives are the control of individuals over their personal data as well as on simplifying the regulatory environment for international organizations by unifying the regulation within the EU \cite{GDPR2016}. 

With regard to privacy, one important requirement is that any processing of personal data must be justified. To that end, simplified speaking, any data that can be linked to an individual has to be regarded as personal data. Furthermore, in order to process data, legal grounds or individual consent reflect a justification. 
According to a ruling of the European Court of Justice, a video stream containing faces corresponds to containing personal data, and, accordingly, underlies the GDPR regulation \cite{EuGH2014}. 
If, however, a video stream shows no personal data, the GDPR does not apply and no consent or legal basis is required in that regard. 

One common approach to remove personal data from any data to prevent the application of the GDPR is its anonymization, i.e., the removal of any link to an individual person. Once anonymized, the data does no longer link to an individual person and does not underlie the GDPR regulation. 

At the same time, however, the general interest and the concerns to privacy of an individual need to be traded-off. In this case, the general interest is usually considered as high, whereas the concerns to privacy are considered low, since the processing merely serves the purpose of allowing processing of data whilst ensuring privacy on an individual level. Accordingly, no consent is required. To that end, one option to approach this task is to perform anonymization on the device itself. Thus, no unanonymized data is available to any party and that the unanonymized data is not stored. In this case, anonymization is be performed \textit{on edge}.


Therefore, to conclude, if the raw data leaves the camera, it underlies the GDPR and a consent or legal basis is required for its processing. On the other hand, if data remains on the device or the transferred data is anonymized, it does not underlie the GDPR regulations. This understanding serves as a design guidance and will be picked up later again.

In the following, we explore technical approaches that preserve privacy in video-based surveillance systems.

\subsection{Approaches to Preserve Privacy in Video-based Surveillance}

Most approaches to ensure privacy in video surveillance are based on anonymizing the video feed, i.e. removing any data that may reveal an identity. To that end, a two-step process is usually deployed: First, identity revealing image segments (e.g. faces) are identified and, second, modified. 
Whilst we acknowledge that there are many identity-revealing segments within an image, we focus on faces throughout this work. Thus, the above depicted two-step process corresponds to face recognition and its anonymization.

The task of face recognition has been widely addressed in research. Indeed, especially since the rise of deep learning, face recognition is increasingly addressed through neural networks. Publications addressing face recognition through deep learning are omnipresent and indicate good results \cite{he2016deep}.

Once faces are identified, commonly-used distortion approaches to ensure privacy are masking, pixelation and blurring \cite{Dufaux2010,Korshunov2014}. 
In masking, the identity-revealing segments of an image are covered with a neutral element, as, for example a \textit{black box} \cite{Newton2005,Moon2009}. Whilst those approaches most certainly address the issue of privacy, they are not applicable to the design challenge of this work, since a black box also removes any data on whether a mask is worn or not.
Second, pixelation refers to the substitution of squared blocks of pixels with its average \cite{Dufaux2010}. Given its simplicity, it is commonly used, as, for example, in television news in order to ensure privacy of individuals within the image. 
Third, in blurring, segments of the picture are blurred, i.e. making the segments less distinct \cite{Schiff2007}. A common approach for blurring is the application of a Gaussian low-pass filter. 

Evidently, the mentioned approaches align with the general concept of \textit{privacy-enhancing technologies (PETs)} \cite{Goldberg1997}. As such, PETs are aimed at protecting an individual's privacy by the use of technical means \cite{Heurix2015}. Indeed, the provision of anonymity, pseudonymity, unlinkability and unobservability of data subjects is a core component of PETs \cite{fischer2001security}. As such, the above described approaches contribute to and reflect PETs. 

Aligned with the idea of PET, Fitwi et al. (2019) \cite{Fitwi2019} propose a lightweight solution to preserve privacy with a special focus on the Internet-of-Things and edge computing. As such, the authors argue that privacy measures should already be built into camera equipment, eventually making the camera a \textit{smart} device only transmitting privacy-preserving video signals. In their approach, the authors rely on pre-trained machine learning models that are loaded onto the camera to compute privacy-preserving video streams on the fly. Similar research is conducted by a number of researchers (e.g., \cite{Streiffer2017,Wang2018}). To that end, those approaches comply with anonymization approaches required to bypass GDPR regulations.

On that note, we could also identify research aimed at preserving privacy that are not suitable for the purpose of this work. Chi and Hu (2015) \cite{Chi2015} and Carillo et al. (2008) \cite{Carrillo2008}, for example, propose algorithmic approaches to encrypt parts of an image that can be decrypted at a later stage in time. Thus, the requirement of irreversible removal of identity-revealing data is not met.
Furthermore, Boyer and Veigl (2015) \cite{Boyer2015} purpose a system that allows for privacy-preserving video surveillance. The authors propose a system that allows access to video surveillance for police investigations. Whilst this use case does underlie the GDPR, the authors introduce an authentication and data protection instance in order to prevent misuse of the video.

\begin{table*}[!htbp]
\centering
\caption{Overview of design requirements, design principles and design features for the proposed privacy-preserving mask recognition artifact}
\label{tab:designoverview}

\begin{tabular}{|l|l|l|}
\hline
\textbf{Design Requirement}                                                                                                                     & \textbf{Design Principle}                                                                                                                                                   & \textbf{Design Feature}                                                                                  \\ \hline
\multirow{2}{*}{\begin{tabular}[c]{@{}l@{}}\textbf{DR1 (Privacy preservation)}: \\ The artifact should not \\ reveal personal data.\end{tabular}} & \begin{tabular}[c]{@{}l@{}}\textbf{DP1a}: Implement a designated\\ privacy preserving service.\end{tabular}                                                                          & \begin{tabular}[c]{@{}l@{}}\textbf{DF1a}: Implement bluring \\ of faces.\end{tabular}                             \\ \cline{2-3} 
                                                                                                                                                & \begin{tabular}[c]{@{}l@{}}\textbf{DP1b}: Do not forward personal \\ data required within the\\  analysis (e.g., raw pictures), \\ but only the result of the analysis.\end{tabular} & \begin{tabular}[c]{@{}l@{}}\textbf{DF1b}: Perform all calculations \\ on edge.\end{tabular}                       \\ \hline
\begin{tabular}[c]{@{}l@{}}\textbf{DR2 (Recognition performance)}: \\ The artifact should reach \\ superior mask recognition \\ performance.\end{tabular}         & \begin{tabular}[c]{@{}l@{}}\textbf{DP2}: Implement an AI-based \\ mask recognition service.\end{tabular}                                                                             & \begin{tabular}[c]{@{}l@{}}\textbf{DF2}: Implement a \\ deep neural network \\ for mask recognition.\end{tabular} \\ \hline
\end{tabular}
\end{table*}

\begin{table*}[!htbp]
\centering
\caption{Overview of identified design options for a privacy-preserving mask recognition artifact}

\label{tab:options}

\begin{adjustbox}{width=1\textwidth}

\begin{tabular}{l|l|l|l|l|l|}
\cline{2-6}
                                        & \textbf{\begin{tabular}[c]{@{}l@{}}Privacy\\ preservation\end{tabular}} & \textbf{\begin{tabular}[c]{@{}l@{}}Mask \\ recognition\end{tabular}} & \textbf{\begin{tabular}[c]{@{}l@{}}Utilized\\ Design Features\end{tabular}} & \textbf{Pro}                                                                 & \textbf{Contra}                                                           \\ \hline
\multicolumn{1}{|l|}{{Baseline}} & None                                                                    & External service                                                   & DF2                                                                         & \begin{tabular}[c]{@{}l@{}}Flexible mask \\ recognition\end{tabular}         & Privacy violations                                                        \\ \hline
\multicolumn{1}{|l|}{\textbf{${Option}_{{centralized}}$}} & On edge                                                                 & External service                                                   & DF1a + DF2                                                                  & \begin{tabular}[c]{@{}l@{}}Non-liftable \\ privacy preservation\end{tabular} & \begin{tabular}[c]{@{}l@{}}Reduced recognition\\ performance\end{tabular} \\ \hline
\multicolumn{1}{|l|}{\textbf{${Option}_{{decentralized}}$}} & not req.                                                                 & On edge                                                            & DF1b + DF2                                                                  & \begin{tabular}[c]{@{}l@{}}Encapsulated \\ functionality\end{tabular}        & \begin{tabular}[c]{@{}l@{}}Fixed mask \\ recognition\end{tabular}         \\ \hline
\end{tabular}

\end{adjustbox}

\end{table*}

In conclusion, this section shows the importance of privacy and how it can be enabled in video-based surveillance. The question on how privacy can be ensured in video-based monitoring systems is well addressed, however, to the best of our knowledge, the intersection of privacy-preservation and mask recognition has not yet been addressed in rigorous research. Precisely, we were not able to find any peer-reviewed work covering the AI-based mask detection in genreal as well as privacy-preserving detectio in specific. Thus, mask detection whilst ensuring privacy constitutes our addressed research gap. 

\section{Artifact Design}
\label{sec:artifactdesign}
In accordance with the design science research  paradigm, we first elaborate on our overview of design choices \cite{meth2015designing} which are depicted in \Cref{tab:designoverview}. As elaborated, we are confronted with two fundamental design requirements (DRs) for the artifact: privacy preservation (DR1) and recognition performance (DR2). 

\begin{figure*}[!htbp]
    \centering
	\includegraphics[width=0.8\linewidth]{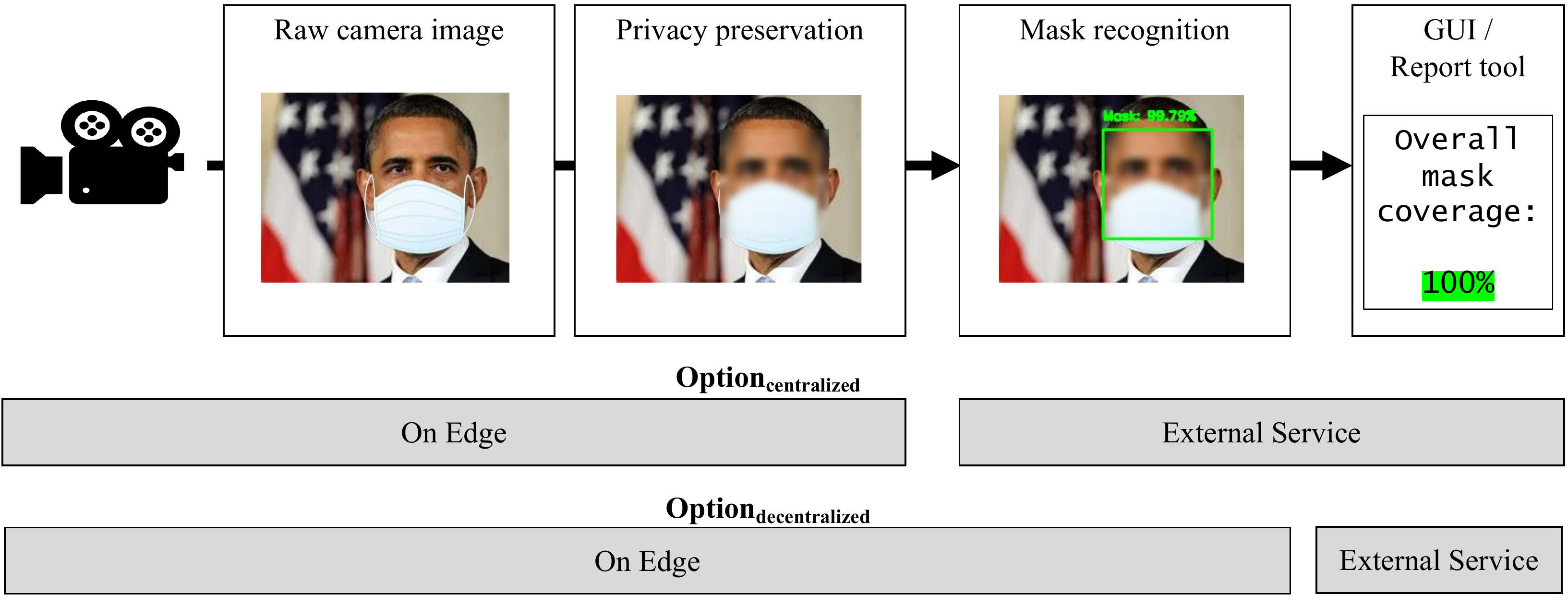}
	\caption{Overview of the overall approach, illustrating the different process steps as well as the two options for deployment at each step.}
	\label{fig:Approach}
\end{figure*}

Regarding DR1, we regard two different design principles (DPs) addressing the removal of personal data of raw video/image data. We can either implement a designated privacy-preserving service locally (e.g. integrated in the camera) which removes any personal data before forwarding the adjusted video data to the potential user / web service (DP1a), or run the analyses locally and only forward the result to the potential user / web service (DP1b). DP1a and DP1b are exclusive. 

For the precise implementations, so-called design features (DFs), we make two choices. For the privacy preserving service (DP1a), we choose to implement a blurring of faces (DF1a). The utilization of pixelation would be another valid option, however Lander et al. \cite{lander2001evaluating} show that if already familiar with a face, blurring provides better anonymization performance. Blurring is state-of-the-art in literature as well as in real world applications as there are multiple privacy friendly configurations \cite{ren2018learning}. 

In the case of not forwarding any personal data, we choose to perform all calculation on edge, i.e., directly on the (camera) hardware---and only output the result of the analysis (DF1b). No personal data is saved on the device and no personal data leaves it. Thus, the GDPR requirements are satisfied. Even if there are malfunctions in the mask recognition service, no violations of GDPR are possible since only a numeric feature on the share of people wearing masks is forwarded.

Regarding DR2, demanding a superior mask recognition performance, we choose to utilize state-of-the art AI techniques (DP2). Precisely, we train a deep neural network to detect masks on images showing people (DF2). Deep neural networks have been proven to achieve close-to-perfect performances in image classification \cite{russakovsky2015imagenet}.


The resulting possible combinations of DFs leaves us with three viable combinations, which are depicted in \Cref{tab:options}. We have to differentiate on two dimensions: which DF is utilized and where the DF is deployed. In terms of deployment, it is either possible to host each of the required services (privacy preservation / AI recognition) on edge / directly on the hardware (``provider side'') or to host the AI recognition service externally (``customer side'')\footnote{Note, the provider is the authority providing the potential sensitive data; the customer is the authority that wants to use the information about mask coverage.}. 
Privacy is violated if there are options for the external / customer side to access personal data from the received information. 

To gain an understanding of the performances of the deep neural net, we start by calculating the \textit{baseline} performance. We require this initial benchmark to later calculate the loss of performance with the raise of privacy. 

${Option}_{{centralized}}$ is a combination of DF1a and DF2. The privacy preservation, in our case blurring of the faces (DF1a), is performed directly on edge \cite{gezer2017extensible} and only the edited images are put forward to an external, centralized service. In this option, the mask recognition (DF2) is deployed externally. The privacy preservation is non-liftable as only the preprocessed data is available to the customer. On the downside, it might lead to worse mask recognition, which we will analyse in the upcoming evaluation. 

${Option}_{{decentralized}}$ is a combination of DF1b and DF2. The privacy preservation is guaranteed as the camera hardware only outputs the numerical results of the on edge mask recognition, e.g. the percentage of people wearing masks. No image data is transmitted. In this option, every aspect is embedded within one encapsulated functionality, i.e. decentralized. On the downside, the artifact does not leave any flexibility, e.g., the option to use the camera output for other analyses. The overview of the options is summarized in \Cref{tab:designoverview} while the resulting overall approach is depicted in \Cref{fig:Approach} on page \pageref{fig:Approach}.

Whilst ${Option}_{{decentralized}}$ most certainly has the highest privacy preservation, it has one major drawback when considering a fleet of cameras: A holistic statement over all cameras is not possible. This can be illustrated using the following example: Suppose there is a fleet of cameras aimed at computing the overall percentage of people wearing a face mask at a train station. Using $\textrm{Option}_{{decentralized}}$, each camera computes the percentage of people wearing a face mask within its video stream and only forwards this percentage to a central server. This server collects the percentage of people wearing a face mask of all cameras and aggregates those to an overall percentage. A single person, however, may appear in multiple video streams. Therefore, the overall percentage of people wearing a face mask may be biased. This aspect could better be accounted for using ${Option}_{centralized}$.
\begin{figure*}[!htbp]
    \centering
	\includegraphics[width=0.8\linewidth]{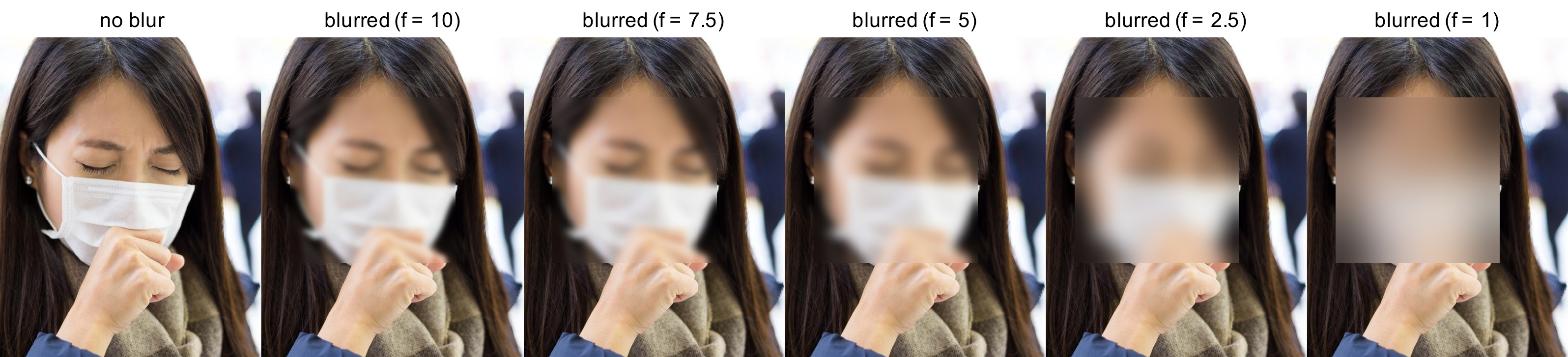}
	\caption{Illustration of blurred faces with different blur factors $f$. A lower $f$ results in higher blurring.}
	\label{fig:blur}
\end{figure*}

\section{Artifact Evaluation}
\label{sec:artifactevaluation}
As a next step, we implement the previously described artifact design and evaluate two aspects, the performance of each option (centralized and decentralized) as well as the influence of the blurring factor, i.e. the ``degree of anonymization'', on the AI's detection performance.

\subsection{Artifact Instantiation}
As described in \Cref{sec:artifactdesign}, there are two possible options for designing a privacy-preserving mask recognition artifact.

${Option}_{centralized}$ ensures in a first step that no personal image data (i.e., facial features which make a person identifiable) is transmitted, e.g. to external web services. Faces are detected on edge (regardless of whether a person is wearing a mask or not) and blurred by applying Gaussian blur. Gaussian blurring is achieved by convolving each pixel of a recognized face with a Gaussian kernel of variable size (factor $f$ indicates the ratio of the kernel size to the image size) in order to create a blurred face. \Cref{fig:blur} on page \pageref{fig:blur} depicts an exemplary face without blurring (left) as well as the same face (second image from left to right) disguised with different blurring factors. While widely used \cite{chandel2013image}, face anonymization with Gaussian blur is discussed controversially, as de-anonymization is possible under certain circumstances \cite{mcpherson2016defeating}; e.g. Dufaux and Ebrahimi (2010) \cite{Dufaux2010} report in the case of applying a Gaussian factor of 8 and while clear (not anonymized) pictures are available to a potential attacker, recognition might be possible.
To counter de-anonymization attempts, it is important to choose a low blur factor (leading to high anonymization, see \Cref{fig:blur}) \cite{gross2009face} as we do in the remainder of this work. Therefore, our proposed artifact uses Gaussian blurring for the moment, but alternatives should be kept in mind.

In a subsequent step, the anonymized image data is processed by a separate externally deployed mask recognition service, which is capable of detecting masks even on anonymized image data. The degree of blurring, and therefore, anonymization, is dicussed in \Cref{sec:tradeoff}.

${Option}_{decentralized}$  performs mask recognition directly on the edge. Thus, a person's privacy is preserved by not passing image data to external services at all, but only transferring aggregated indicators such as the ratio of persons wearing masks to persons without wearing masks. Thus, in this case, anonymizing of image data is not necessary, since images are not transmitted to an external party anyway. From a technical point of view, ${Option}_{decentralized}$ therefore is equivalent to the baseline, where raw data is directly transferred to an externally deployed service, which then takes care of mask recognition without addressing any privacy aspects.

\subsection{Artifact Performance}
\label{sec:artifactperformance}
The evaluation of the artifact is based on image data of persons who either wear a mask or do not---with the aim of being able to distinguish them with the highest possible performance, measured by the metric of \textit{accuracy}. The accuracy indicates the overall proportion of correctly (=true) predicted observations a classifier achieves. It reaches its best value at $1.0$ ($100\%$) and its worst at $0$ ($0\%$).


While there are several publicly available data sets depicting persons' unmasked faces, data sets including persons wearing (medical) masks are rare. Note, as there is no data of mass surveillance footage publicly available, we assume---as existing work shows \cite{ge2017detecting,kh2020using}---that images showing individuals can be retrieved from pictures showing multiple individuals. 

Our evaluation is based on two different data sets showing individuals. The first originates from the machine learning platform Kaggle and contains $1,000$ images, equally shared among masked and unmasked persons \cite{mask_dataset}. Additionally, we utilize a data set of persons from \cite{pyimagesearch} without wearing masks and automatically place an artificial mask in front of 50\% of the persons' faces. An example of an artificially placed mask is displayed in \Cref{fig:Approach}. This artificially created data set contains a total of $686$ images per class. 

To obtain a \textit{baseline} performance, we train a deep convolutional neural network based on the MobileNetV2 architecture \cite{MobileNetv2} and pre-trained it on the ImageNet database \cite{imagenet_cvpr09}, assigning each input image depicting a person's non-blurred face to the classes \emph{mask} or \emph{no mask}. We use MobileNetV2 because on the one hand it is a common architecture for this kind of application and on the other hand it is optimized for edge computing. We also fix the number of epochs at 15 to keep the training time within reasonable limits. For training-test split, we use a common 75/25 ratio.

${Option}_{decentralized}$ is technically equivalent to the base case, at least in terms of modeling, and therefore its performance equals the baseline. Only the kind of deployment and, thus, the aggregation of the model output ensures that privacy is maintained in comparison to the baseline.

For ${Option}_{centralized}$, however, we use images with blurred faces to train a deep neural network with the same architecture as in the {\it base case}. To blur the faces we apply the state-of-the-art dlib face recognition model based on the ResNet architecture \cite{he2016deep} combined with a Gaussian filter which is applied to the rectangular facial section identified by the recognition model.

\begin{table}[!htbp]
\centering
\caption{AI performance comparison of the two regarded options with a blurring factor of $f = 5.0$}
\label{tab:performance}

\begin{adjustbox}{width=1\linewidth}

\begin{tabular}{|l|l|l|}
\hline
                   & \multicolumn{2}{c|}{\textbf{Accuracy}} \\ \cline{2-3} 
                   & \textbf{Real Data}   & \textbf{Artificial Data} \\ \hline
Baseline           & 1.00        & 0.99            \\ \hline
\textbf{${Option}_{centralized}$}\footnotemark & 0.98\small{\quad($-$2\%)} & 0.95\small{\quad($-$4\%)}     \\ \hline
\textbf{${Option}_{decentralized}$}           & 1.00\small{\quad($\pm{0}\%$)} & 0.99\small{\quad($\pm{0}\%$)}     \\ \hline
\end{tabular}

\end{adjustbox}

\end{table}

\Cref{tab:performance} compares the performances. 
The accuracy of ${Option}_{centralized}$ decreases only a few percent on both data sets when using a blur factor of 5.  Overall, the performance loss between {\it baseline} \ ${Option}_{decentralized}$ and ${Option}_{centralized}$ is therefore only minimal. This implies that an increase in privacy (due to the blurring factor 5) causes only a small loss in model performance.

\subsection{Robustness Check}
\label{sec:sanity}
The aim of the paper is to show that mask recognition and privacy preservation do not contradict each other. For this purpose we show the overall feasibility and the difference between two different options, but do not put a strong focus on tuning the models used. In the absence of data sets with self-sewn, colored or printed everyday masks as they are worn quite often, we additionally validate a model trained on both data sets (see \Cref{sec:artifactperformance}) on a small number of self-recorded images. We collect 50 images of people wearing colored masks to provide a basis for demonstrating that our proposed approach is generalizable and that everyday masks can be reliably recognized.

The results show that an accuracy of 90.32\% is achieved with a blurring factor of 5, although we have tried to select masks that are intentionally difficult to detect (e.g., painted mouth or skin-colored). However, with a higher blurring factor, the accuracy of the model becomes progressively worse, especially due to  a drastically decreasing rate of true positives.

\subsection{Privacy vs. Performance Trade-off}
\label{sec:tradeoff}

Results illustrated in the previous subsection show that instant mask recognition ({\it baseline} and ${Option}_{decentralized}$) performs slightly better than anonymizing faces in advance (${Option}_{centralized}$). Even if these differences are still very small at a blurring factor of 5 (as depicted in \Cref{tab:performance}), it can be assumed that the performance of mask recognition decreases with increasing blur.

Therefore, we have trained the mask recognition model on faces with varying degrees of blurring (see \Cref{fig:blur}). The degree of privacy preservation corresponds to the inverted blur factor. This assumption is based on the underlying mechanism of the blurring factor itself. At very large values and thus a very small kernel size, this factor causes an almost non-blurred face. In contrast, a factor of $1$ corresponds to a kernel size that is equal to the image size, causing a maximum blurred face.

\begin{figure}[htbp]
    \centering
	\includegraphics[width=1\linewidth]{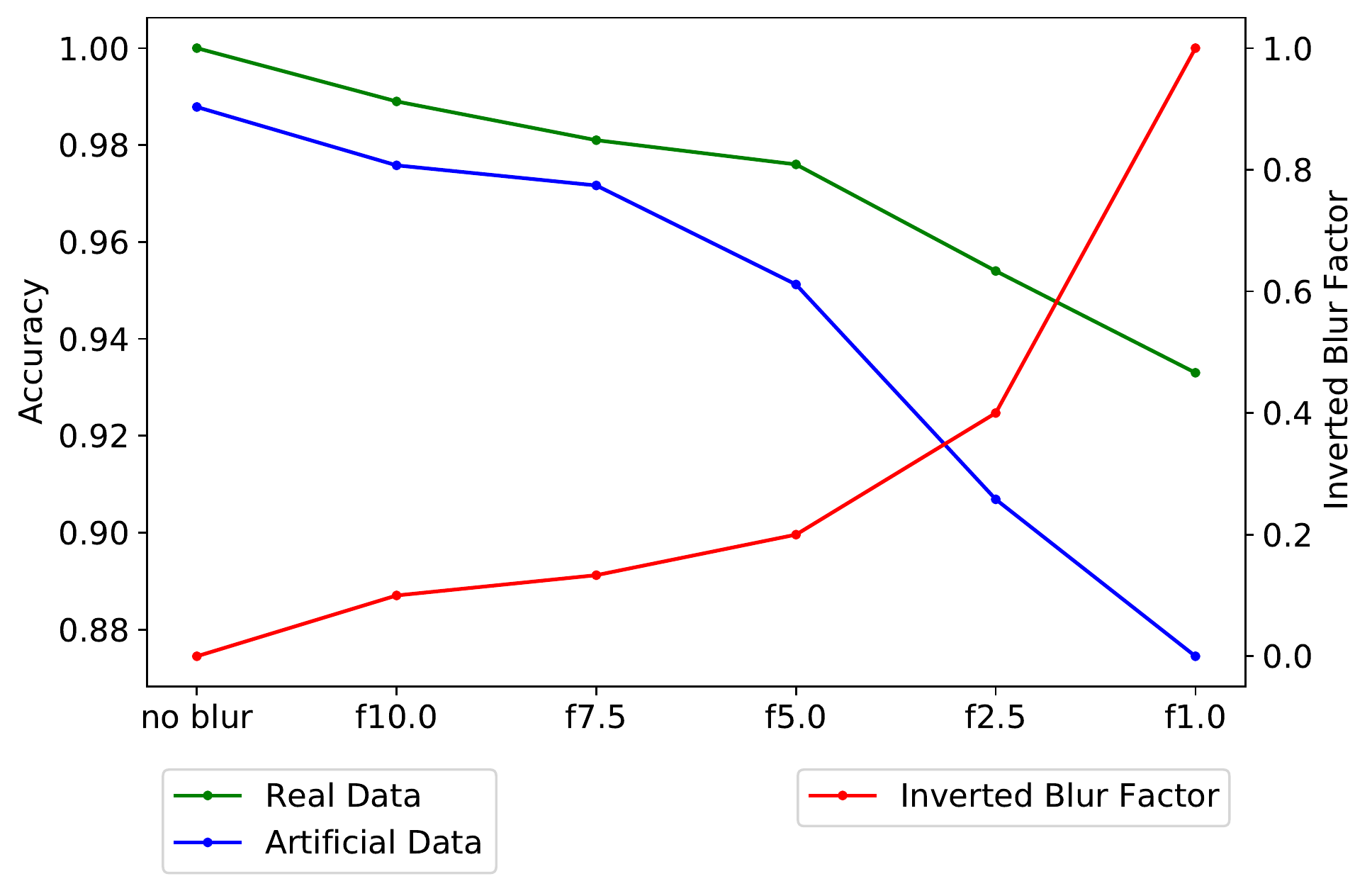}
	\caption{Empirical results illustrating the trade-off between privacy-preservation and AI performance with the two implemented data sets}
	\label{fig:Tradeoff}
\end{figure}

\begin{figure}[htbp]
    \centering
	\includegraphics[width=1\linewidth]{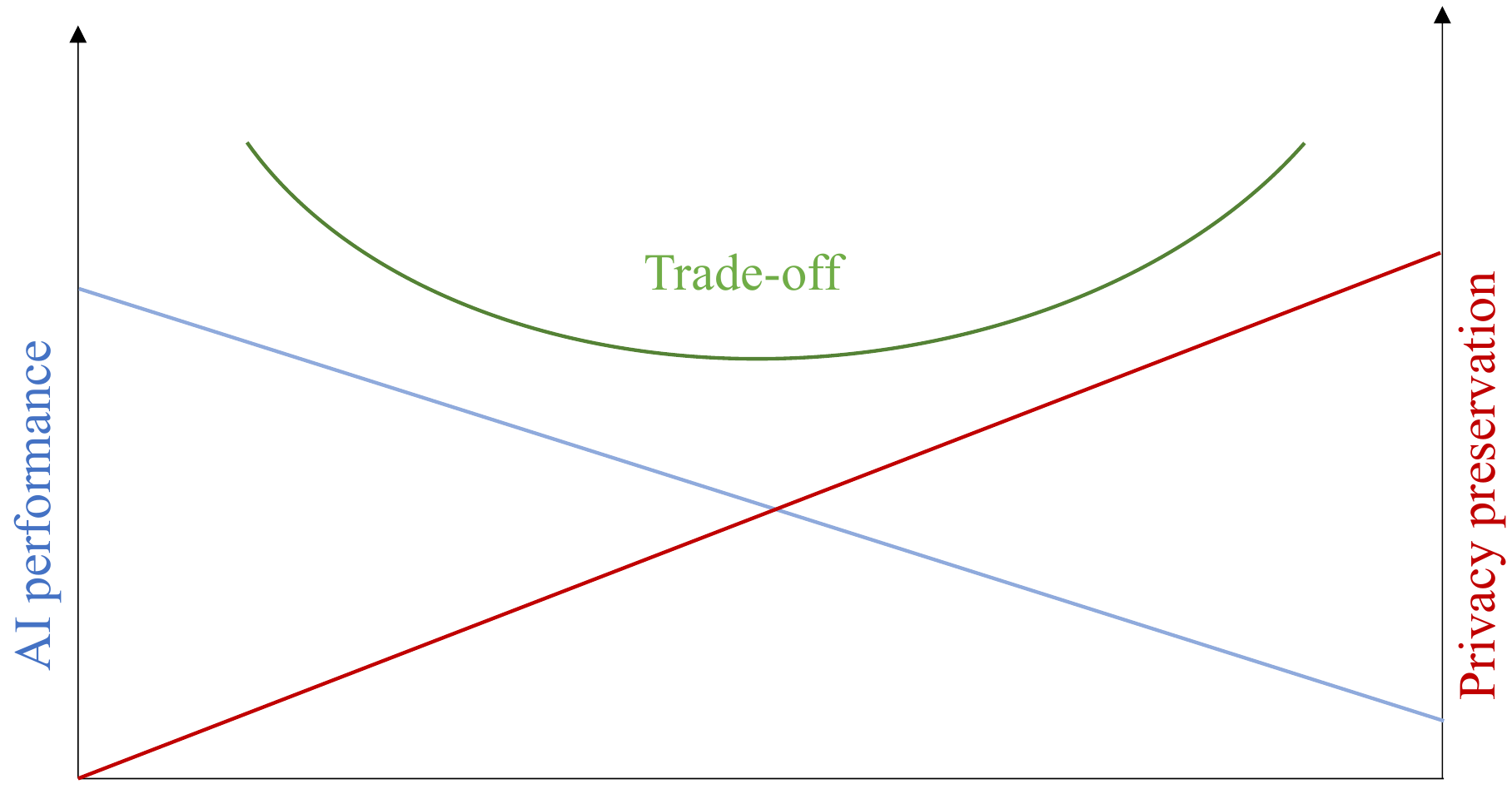}
	\caption{Hypothesized relationship between AI performance and privacy preservation}
	\label{fig:Tradeoff_abstract}
\end{figure}

\Cref{fig:Tradeoff} shows how the model performance of mask recognition actually decreases on both data sets with increasing blur. The more a face is disguised and, thus, the higher the privacy preservation, the more the model performance decreases. We call this loss of performance ``price of privacy''. We theorize this observation could be generalized into the tradeoff between AI performance and privacy preservation as qualitatively illustrated in \Cref{fig:Tradeoff_abstract}. In our case, however, despite the maximum possible blur ($f = 1$), it is still within an acceptable range of 6\% to 11\% loss of model accuracy depending on the data set.

\section{Discussion}
\label{sec:discussion}

Based on the promising results we include a statement of the broader impact of our work, including confounding factors, the artifact's potential ethical aspects and future societal consequences. 

If we imagine an ${Option}_{centralized}$-based system being in place on a large scale, possible benefits and disadvantages arise from our research. On the upside, we would be able to allow a recognition of masks coverage for research purpose, e.g. in order to monitor whether people reduce infection risks. As, for this option, we propose a removal of personal data on edge, it is not possible for consumers of the service to trace back individuals---as the raw image is not accessible. However, we need to discuss two types of errors: First, the privacy preservation service failing and, second, the mask detection service failing.

Although we were able to show accuracy rates are fairly high throughout our experiments, no system is perfect. The question of the data quality, in this case camera resolution, positioning angle and ability to capture a space holistically, are essential for our approach. As we only used data of high quality, results might be inferior with other data. Furthermore, we did not test it within a real scenario, but on pictures only showing single persons---the segmentation of large crowds would need to be performed by a different service (on edge, as suggested by Wang et al. (2020) \cite{kh2020using}). In any case, if the system would fail, it might be possible to trace back individuals, as there is no guarantee of correct face blurring. Reasons for failure could be many, e.g., technical errors in the camera image, but also biases due to training. The latter could lead to discrimination of certain groups which were not included in the initial training.

If the mask recognition service fails, different problems might be the result. On the one hand, if the service detects too many masks being worn (when in fact less are actually worn), situations might be classified as ``safe'' by the system, when they are, in fact, not safe. On the other hand, when the system does not detect all masks (although they are worn) it could issue wrong alarms, or, depending on the ability of the system and the severity of the alarm, lock down areas.

We, therefore, encourage further research in the reliance and fairness of AI-based systems, especially in the area of surveillance. For instance, as previous work has shown, the COMPAS system used by US courts to assess defendants' risk of recidivism, was unfair towards black people---although it was in productive use \cite{angwin2016machine}.

In addition to those errors, we also see a confounding factor in case one person appears in multiple video streams---as already mentioned in Section \ref{sec:artifactdesign}. In detail, each video stream provides a sample of the overall population and one person within the population may appear in multiple samples (i.e. video streams). Overall, this results in a prediction bias. If ${Option}_{decentralized}$ is applied, no association between multiple video streams are possible and less accurate results due to the bias have to be accepted. If ${Option}_{centralized}$ is applied, however, one may add an additional process step aimed at identifying the same person (yet, still anonymous) on multiple video streams, as, for example, through the clothes they are wearing. Without further investigation, we believe that an additional processing may reduce, but will not completely remove the risk of less accurate results.
\section{Conclusion}
\label{sec:conclusion}
In 2020, the COVID-19 disease has evolved into a global pandemic---and governments all over the world reacted by taking different measures, so-called non-pharmaceutical interventions (NPIs). One of these NPIs, in line with current research \cite{greenhalgh2020face}, constitutes the obligation to wear masks in public spaces like stores, restaurants, etc. \cite{aljazeera2020}. One option to monitor the compliance of citizens wearing masks is to rely on video surveillance, e.g., closed circuit television (CCTV). This option would have the benefit of being fully scalable and possibly automatable. However, the General Data Protection Regulation (GDPR) has strong requirements on the handling of personal data (including clearly visible faces) \cite{GDPR2016,EuGH2014} and citizens might feel a violation of their privacy when being surveillanced \cite{stuart2017beyond}. 

To address this gap, this work proposes an AI-based surveillance artifact which ensures both (a) privacy and (b) high performance of mask recognition. We demonstrate different options of privacy-preserving methods (in-line with GDPR) and their resulting performances. Depending on the chosen option, e.g., on edge or as an external service, our results show accuracies between 95\% and 99\%. In conclusion, we show that privacy-preserving mask recognition is well-feasible.

By designing, implementing and evaluating our artifact we contribute to the body of knowledge in three meaningful ways. First, our novel artifact can be utilized to allow for a privacy-preserving monitoring of mask coverage during an epidemic like the flu or a pandemic like COVID-19. Second, we evaluate different design choices on how to build the artifact and elaborate on their performances and capabilities to preserve privacy. Finally, we theorize on the trade-off between privacy preservation and AI performance---as AI performance decreases with increased privacy preservation and vice-versa. Our findings indicate that even with the highest degree of privacy preservation we applied, the loss of AI performance does not exceed 11\%.

The generalizability of these results is subject to certain limitations. One shortcoming of our work is the fact that we utilize only images containing single faces. To address this issue, recent work  examines the possibility to pre-process larger images to extract single images with one person per image, so-called segmentation \cite{ge2017detecting,kh2020using}. This step could be easily implemented on-edge as well. Additionally, under certain circumstances recent research shows the possibility to de-anonymize blurred pictures \cite{mcpherson2016defeating}. To counter that, we use high blurring in our presented ${Option}_{centralized}$, while ${Option}_{decentralized}$ does not require blurring at all. 

Apart from these limitations, further research should be undertaken to investigate other aspects of the endeavor. Engaging in a dialogue with hardware providers on the (technical) possibilities of implementing privacy-preserving measures would be worthwhile. Especially in our presented ${Option}_{centralized}$, where only the blurring of faces occurs on chip the options for other applications (apart from mask recognition) are manifold. As there is no traceable personal data left as the output of the camera system, the video stream could be utilized for other cases. For instance, following the introductory example \cite{bbc2020}, the systems could additionally be used to monitor distance rules, count the total amount of people, etc. In any of these cases, a convenient user interface would need to be designed, which we did not address in this work. Depending on the use case, it would be important to build user centric interfaces for end users to ensure the technology can be put to practice. In regards to the case presented in this work, mask recognition, different applications are possible; it could be utilized to simply count how many citizens are compliant (e.g., for research purposes) or as a 
``red alert'' warning system, i.e., within smaller and crowded places. In the latter, it would need to be discussed which actions would be undertaken as a countermeasures if people do not comply, e.g., automatic announcements over speakers or else. Future work will hopefully give insights into these suggestions, as a promising field of research lies ahead.

\section*{Acknowledgements}
The research reported in this paper has furthermore been supported by the German Federal Ministry of Education and Research within the framework of the project KASTEL\_SKI in the Competence Center for Applied Security Technology (KASTEL).

\bibliographystyle{ieeetr}
\bibliography{references}
\end{document}